\documentstyle[aps,twocolumn,graphicx,floats,prl,xspace,amsmath]{revtex}

\newcommand{\micron}{$\mu$m\xspace}

\begin{document}

\wideabs{
\title{Hydrodynamic Coupling of Two Brownian Spheres to
a Planar Surface}

\author{Eric R. Dufresne$^{(1)}$, Todd M. Squires$^{(2)}$, 
  Michael P. Brenner$^{(3)}$, and David G. Grier$^{(1)}$}

\address{$^{(1)}$Dept. of Physics, James Franck Institute, and
Institute for Biophysical Dynamics\\
The University of Chicago, Chicago, IL 60637\\
$^{(2)}$ Dept. of Physics, Harvard University, Cambridge, MA 02138\\
$^{(3)}$Dept. of Mathematics, Massachusetts Institute of Technology, 
Cambridge, MA 02139}

\date{\today}

\maketitle

\begin{abstract}
  We describe direct imaging measurements of the collective and relative 
  diffusion of
  two colloidal spheres near a flat plate.
  The bounding surface modifies the spheres' dynamics, even at
  separations of tens of radii.
  This behavior is captured by a stokeslet analysis of fluid flow
  driven by the spheres' and wall's no-slip boundary conditions.
  In particular, this analysis reveals surprising asymmetry in
  the normal modes for pair diffusion near a flat surface.
\end{abstract}


} 

Despite considerable progress over the past two centuries \cite{vandeven89}
hydrodynamic properties of all but the simplest colloidal systems
remain controversial or unexplained.
For example, velocity fluctuations in sedimenting 
colloidal suspensions
are predicted to diverge with system size \cite{caflisch85}.
Experimental observations indicate, on the other hand, that long-wavelength 
fluctuations
are suppressed by an as-yet undiscovered mechanism 
\cite{xue92,nicolai95,segre97}.
One possible explanation is that hydrodynamic coupling to bounding surfaces
may influence particles' motions to a greater extent
and over a longer range than previously suspected 
\cite{brenner99}.
Such considerations invite a renewed examination of how hydrodynamic
coupling to bounding surfaces influences colloidal particles'
dynamics.

This Letter describes an experimental and theoretical investigation
of two colloidal spheres' diffusion near a flat plate.
Related studies have addressed the dynamics of
two spheres far from bounding walls \cite{crocker97,meiners99},
and of a single sphere in the presence of one or two walls
\cite{faucheux94}.
Confinement by two walls poses particular difficulties
since available theoretical predictions apply only for highly
symmetric arrangements \cite{happel91}, or else 
contradict each other \cite{liron76,lobry96}.
The geometry we have chosen avoids some of this complexity
while still highlighting the range of non-additive hydrodynamic coupling in
a many-surface system.

We combined  optical tweezer manipulation \cite{grier97} 
and digital video microscopy \cite{crocker96}
to measure four components of the pair diffusion
tensor for two colloidal spheres
as a function of their center-to-center separation $r$ and of their
height $h$ above a planar glass surface.
Measurements were performed on silica sphere of radius 
$0.495 \pm 0.025$~\micron (Duke Scientific lot 21024)
dispersed in a layer of water $140 \pm 2$~\micron thick.
The suspension was
sandwiched between a microscope slide and a \#1 coverslip whose surfaces
were stringently cleaned before assembly \cite{hair70} and
whose edges were hermetically sealed with a
uv cured epoxy (Norland type 88) to prevent evaporation and 
suppress bulk fluid flow.
A transparent thin film heater bonded to the microscope slide
and driven by a Lakeshore LM-330
temperature controller maintained the sample volume's temperature
at $T = 29.00\pm 0.01^\circ$C, as measured by a platinum
resistance thermometer.
The addition of 2~mM of NaCl to the solution
minimized electrostatic interactions
among the weakly charged spheres and glass surfaces by reducing the
Debye screening length to 2~nm.
Under these conditions, the individual spheres'
free self-diffusion coefficients are expected to be
$D_0 = k_B T/(6 \pi \eta a) = 0.550 \pm 0.028~\mathrm{\mu m^2/sec}$,
where $\eta = 0.817$~cP is the electrolyte's viscosity \cite{viscosity}.

\begin{figure}[htbp]
  \centering
  \includegraphics[width=3in]{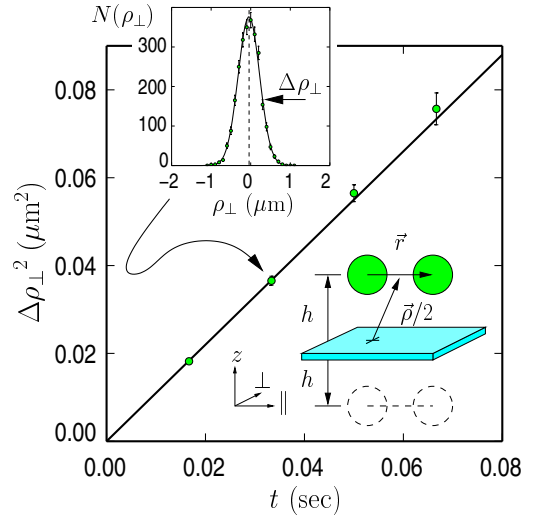}
  \vspace{1ex}
  \caption{Measurement of a typical pair diffusion coefficient in the
    geometry depicted in the lower inset.  Dashed spheres represent
    hydrodynamic images.
    The upper inset shows the histogram of two spheres' collective
    displacements in the $\perp$ direction starting from
    $h = 25.5\pm 0.7$~\micron,
    $r = 7.00 \pm 0.25$~\micron, and $\rho = 0$, 
    after free diffusion for $\tau = 1/30$~sec.   
    Fitting to a Gaussian yields the rms displacement
    $\Delta \rho_\perp(r,h,\tau)$.
    The main plot tracks the evolution of $\Delta \rho_\perp^2(r,h,\tau)$
    together with a least squares fit to
    Eq.~(\protect\ref{eq:stokeseinstein}) for the diffusion coefficient
    $D^C_\perp(r,h)$.  The result is indicated by an arrow in
    Fig.~\protect\ref{fig:data}(d).}
  \label{fig:diffusion}
\end{figure}

The spheres' motions were tracked with 
an Olympus IMT-2 optical microscope using a
100$\times$ NA 1.4 oil immersion objective.
Images acquired with an NEC TI-324A CCD camera were recorded on
a JVC-822DXU SVHS video deck before being digitized with a
Mutech MV-1350 frame grabber at 1/60~sec intervals.
Field-accurate digitization was assured by interpreting the
vertical interlace time code recorded onto each video field.
The spheres' locations $\vec r_1(t)$ and $\vec r_2(t)$
in the image acquired at time $t$ then were measured to within 20~nm 
using a computerized centroid tracking algorithm \cite{crocker96}.

A pair of spheres was placed reproducibly in a plane parallel
to the glass surfaces using optical tweezers \cite{grier97}.
These optical traps were created
with a solid state laser (Coherent Verdi)
whose beam was brought to a focus within the sample volume
by the microscope's objective.
Resulting optical gradient forces suffice to
localize a silica sphere at the focal point despite
random thermal forces \cite{grier97}.
Two optical traps were created by alternating the
focused laser spot between two positions
in the focal plane at 200~Hz using a galvanometer-driven mirror
\cite{sasaki91}.
Diverting the trapping laser onto a beam block every few cycles
freed the spheres to diffuse away from this well defined initial
condition.
Resuming the trap's oscillation between the two trapping points
resets the spheres' positions.
Alternately trapping and releasing the spheres allowed us to 
sample their dynamics efficiently in a particular geometry.
Allowing the spheres only $\tau = 83$~msec (5 video fields) of freedom
before retrapping them for 16~msec (less than 1 video field)
ensured that their
out-of-plane motions, $\Delta z < \sqrt{2D_0\tau} = 0.4$~\micron,
cause negligible tracking errors.

Because optical tweezers form in the microscope's focal plane, their
height $h$ relative to the coverslip's surface can be adjusted
from 1 to 30~\micron with
0.5~\micron accuracy by adjusting the microscope's focus.
For a given height, we continuously varied
the spheres' initial separation between
2~\micron and 10~\micron at 0.025~Hz for a total of 20~minutes.
This procedure yielded 60,000 samples of the spheres' dynamics 
in 1/60~sec intervals divided into sequences 5/60~sec long
for each value of $h$.
These trajectory data were decomposed into 
cooperative motions $\vec \rho = \vec r_1 + \vec r_2$ and
relative motions $\vec r = \vec r_1 - \vec r_2$
either
perpendicular or parallel to the initial separation vector, and
binned according to the initial separation, $r$.
The diffusion coefficients $D_\psi(r,h)$ associated with each
mode of motion $\psi(r,h,\tau)$ at each height and 
initial separation were then obtained from
the Stokes-Einstein formula
\begin{equation}
  \langle \Delta \psi^2(\tau) \rangle = 2 D_\psi(r,h) \tau,
  \label{eq:stokeseinstein}
\end{equation}
where the angle brackets indicate an ensemble average.

\begin{figure}[htbp]
  \centering
  \includegraphics[width=3in]{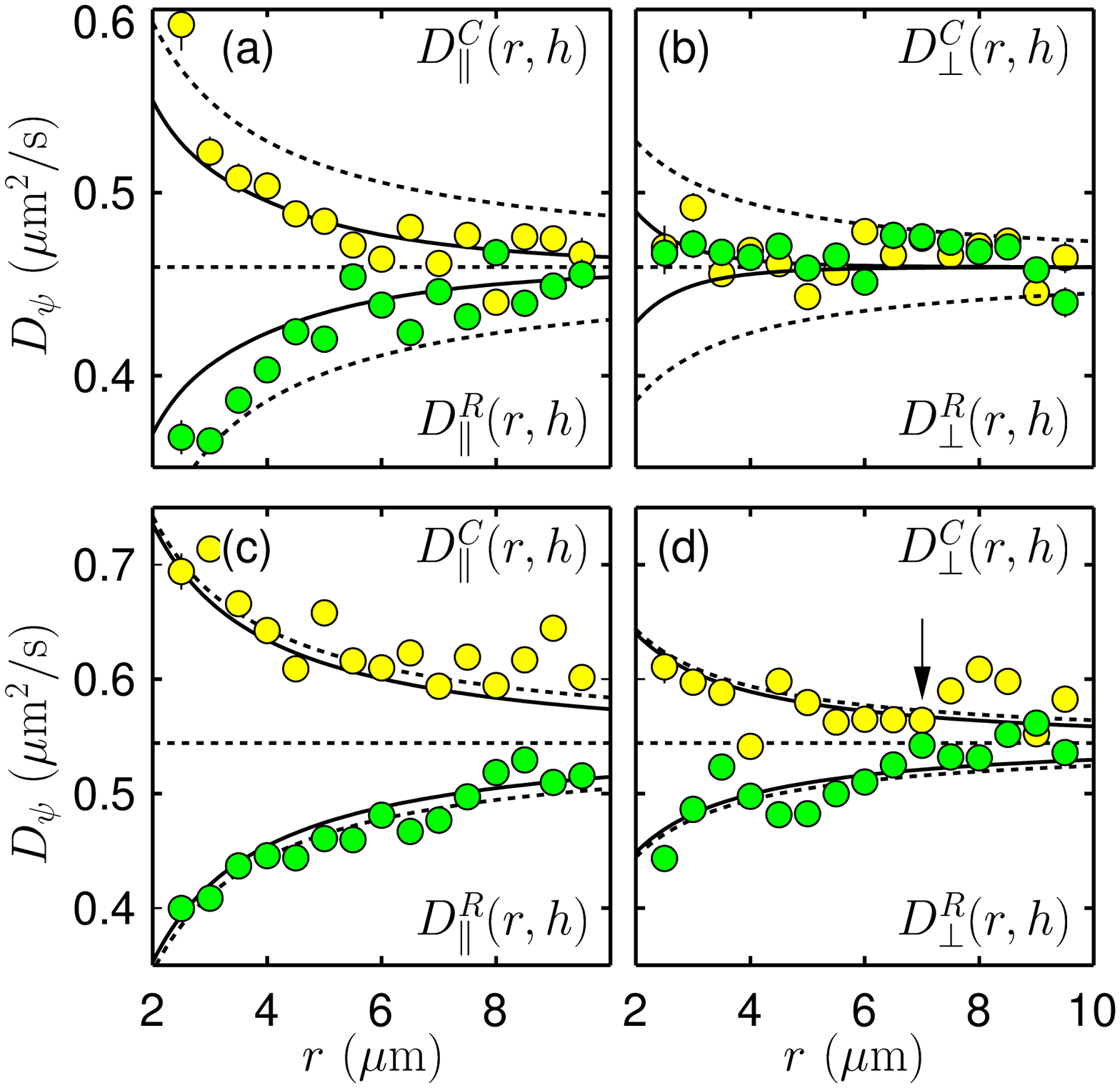}    
  \vspace{1ex}
  \caption{Pair diffusion coefficients for
    1~\micron diameter silica spheres as a function of center-to-center
    separation $r$ and at two different center-to-surface
    heights $h$.  
    (a) and (b): $h = 1.55 \pm 0.66$~\micron.
    (c) and (d): $h = 25.5 \pm 0.7$~\micron.
    (a) and (c): Collective and relative motions parallel to the
    initial separation vector.
    (b) and (d): Perpendicular.
    Dashed curves result from linear superposition of drag coefficients.
    Solid curves result from the theory described here, with no adjustable
    parameters.
    Horizontal dashed lines indicate the asymptotic self-diffusion coefficient
    $D_{xy}(h)$ from Eq.~(\protect\ref{eq:faxenb}).
    }
  \label{fig:data}
\end{figure}

Fig.~\ref{fig:diffusion} shows typical data for one mode of motion
at one height and starting separation.
Diffusion coefficients $D_\psi(r,h)$ extracted from
least squares fits to Eq.~(\ref{eq:stokeseinstein}) appear
in Fig.~\ref{fig:data}
as functions of $r$ for the smallest and largest accessible values of $h$.
The horizontal dashed lines in Fig.~\ref{fig:data} indicate
the spheres' asymptotic self-diffusion coefficients.
Measurements' deviations from these limiting values reveal the influence of 
the spheres' hydrodynamic interactions.

Particles moving through a fluid at low Reynolds number excite
large-scale flows through the no-slip boundary condition at
their surfaces. 
These flows couple distant particles' motions, 
so that each particle's
dynamics depends on the particular configuration of the entire collection.
This dependence is readily calculated using Batchelor's
generalization of Einstein's
classic argument \cite{batchelor76}:
The probability to find $N$ particles at equilibrium
in a particular configuration
$\{\vec r_1, \dots,\vec r_N\}$
depends on their interaction
$\Phi(\vec r_1, \dots , \vec r_N)$
through Boltzmann's distribution,
$P(\vec r_1, \dots , \vec r_N) = \exp \left[ - \Phi/(k_B T) \right]$.
The corresponding force
$-\nabla \Phi = k_B T \, \nabla P/P$
drives a probability flux
$ k_B T \, {\bf b} \, \nabla P $,
where ${\bf b} ( \vec r_1, \dots, \vec r_N)$ is the particles' mobility tensor.
The system reaches equilibrium when this interaction-driven flux
is balanced by a diffusive flux $- {\bf D} \nabla P$.
It follows that the $N$-particle
diffusivity is ${\bf D} =  k_B T \, {\bf b}$.  
Elements of ${\bf D}$ lead to generalized Stokes-Einstein
relations \cite{ermak78}
\begin{equation}
  \langle \Delta r_{i\alpha}(\tau) \Delta r_{j\beta}(\tau)\rangle = 
  2D_{i\alpha,j\beta} \, \tau.
\end{equation}
describing how particle $i$'s motion in the $\alpha$ direction
couples to particle $j$'s in the $\beta$ direction.

The mobility tensor for spheres of radius $a$ has the form
\begin{equation}
  b_{i\alpha,j\beta} = \frac{\delta_{i\alpha,j\beta}}{6 \pi \eta a} +
  b^e_{i\alpha,j\beta}.
\end{equation}
$b^e_{i\alpha,j\beta}$ is the Green's function
for the flow at $\vec r_i$ in the $\alpha$ direction due to
an external force at $\vec r_j$ in the $beta$ direction.
In the present discussion, it accounts for no-slip boundary
conditions at all other surfaces in the system.

If the spheres are well separated, we may approximate
the flow field around a given sphere by a stokeslet,
the flow due a point force at the sphere's location.
This approximation is valid to leading order in the spheres'
radius.
The Green's function for the flow at $\vec x$ in the $\alpha$ direction
due to a stokeslet at $\vec r_j$ in the $\beta$ direction is 
\cite{pozrikidis92}
\begin{equation}
  G^S_{\alpha\beta}(\vec x - \vec r_j) = \frac{1}{8\pi\eta}
  \left[\frac{\delta_{\alpha\beta}}{|\vec x - \vec r_j|} +
    \frac{(\vec x - \vec r_j)_\alpha(\vec x - \vec r_j)_\beta}{
      |\vec x - \vec r_j|^3} \right]
\end{equation}
so that $b^e_{i\alpha,j\beta} = G^S_{\alpha\beta}(\vec r_i - \vec r_j)$.
In the particular case of two identical spheres,
diagonalizing the resulting diffusivity tensor ${\bf D}$ yields 
the diffusion coefficients for two collective (C) modes
and two relative (R) modes along directions perpendicular ($\perp$)
and parallel ($\|$) to the initial separation
\cite{batchelor76}
\begin{eqnarray}
  \frac{D^{C,R}_\perp (r)}{D_0} & = & 1 \pm \frac{3}{4} \, \frac{a}{r} + 
    {\cal O}\left( \frac{a^3}{r^3} \right) \label{eq:a} \\
  \frac{D^{C,R}_\| (r)}{D_0} & = & 1 \pm \frac{3}{2} \, \frac{a}{r} +
    {\cal O}\left( \frac{a^3}{r^3} \right), \label{eq:b}
\end{eqnarray}
where the positive corrections apply to collective modes
and the negative to relative.
The collective diffusion coefficients $D^C_\perp$ and $D^C_\|$ 
are enhanced by hydrodynamic coupling
because fluid displaced 
by one sphere entrains the other.
Relative diffusion coefficients $D^R_\perp$ and $D^R_\|$
are suppressed, on the other hand, by the need to
transport fluid into and out of the space between the spheres.
 
Introducing a planar boundary into this system adds considerable complexity.
The flow field around a small sphere located a
height $h$ above a horizontal wall 
is most easily calculated by the method of images \cite{blake71}, 
in which the wall's no-slip boundary condition is satisfied by 
placing a stokeslet (S), a source doublet (D), and a stokeslet doublet (SD)
a distance $h$ below the plane of the wall \cite{blake71,pozrikidis92}.
The flow due to this image system is
described by the Green's function
\begin{multline}
  G^W_{\alpha\beta}(\vec x - \vec R_j) = -G^S_{\alpha \beta}(\vec x - \vec R_j) \\
  + 2h^2 G^D_{\alpha\beta}(\vec x - \vec R_j)
  - 2h G^{SD}_{\alpha\beta}(\vec x - \vec R_j)
\end{multline}
where $\vec R_j = \vec r_j - 2h \hat z$ is the position of sphere $j$'s image, and
\begin{eqnarray}
  G^D_{\alpha\beta}(\vec y) & = & (1 - 2 \delta_{\alpha z}) \, 
  \frac{\partial}{\partial y_\beta} \left( \frac{y_\alpha}{y^3}\right) \\
  G^{SD}_{\alpha\beta}(\vec y) & = & (1 - 2 \delta_{\alpha z}) \,
  \frac{\partial}{\partial y_\beta}G^S_{\alpha z}(\vec y)
\end{eqnarray}
are Green's functions for a source dipole and a stokeslet doublet, respectively.
The flow field set up by the
image system (and thus by the wall's no-slip boundary condition) 
entrains the sphere through
$b^e_{i\alpha,i\beta} = G^W_{\alpha\beta}(\vec r_i - \vec R_i)$
and decreases its mobility.
Two independent modes emerge from this analysis, 
one ($z$) normal to the wall and the
other ($xy$) parallel, with diffusivities \cite{happel91}
\begin{eqnarray}
  \frac{D_z (h)}{D_0} & = & 1 - \frac{9}{8} \, \frac{a}{h}
    + {\cal O} \left(\frac{a^3}{h^3}\right) \label{eq:faxena}\\
  \frac{D_{xy} (h)}{D_0} & = & 1 - \frac{9}{16} \, \frac{a}{h}
    + {\cal O} \left(\frac{a^3}{h^3}\right). \label{eq:faxenb}
\end{eqnarray}

Eqs.~(\ref{eq:a}) and (\ref{eq:b}) should suffice for two spheres
far from bounding surfaces.
Similarly, the spheres' motions should decouple when the influence of
a nearby wall dominates; Eqs.~(\ref{eq:faxena}) and (\ref{eq:faxenb})
then should  apply.
At intermediate separations, however, neither set of formulas
is accurate.
Naively adding the drag coefficients due to sphere-sphere
and sphere-wall interactions yields
$D_\psi^{-1}(r,h) = D_\psi^{-1}(r) + D_{xy}^{-1}(h) - D_0^{-1}$.
Results of this linear superposition approximation appear as
dashed curves in Fig.~\ref{fig:data}.
While adequate for spheres more than 50 radii from the wall
[(c) and (d)], linear
superposition underestimates the wall's influence
for smaller separations [(a) and (b)].

A more complete treatment not only resolves these quantitative
discrepancies but also reveals an additional surprising
influence of the bounding surface on the spheres' dynamics: the 
highly symmetric and experimentally accessible modes parallel to
the wall are no longer independent.

\begin{figure}[htbp]
  \centering
  \includegraphics[width=3in]{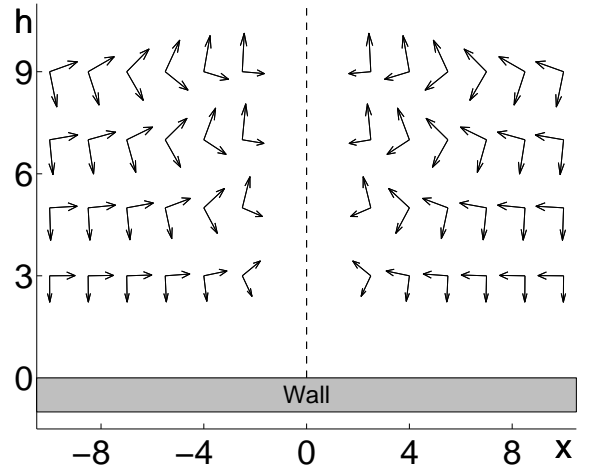}    
  \caption{Cross-sectional view of the diffusive modes for two 
    spheres near a wall.
    Collective motion normal to the wall
    becomes increasingly coupled with relative motion parallel to the wall
    as $h$ approaches $r$.  Collective normal modes at large $r$
    cross over continuously to relative parallel modes as $r$ decreases.
    The dashed line at $x=0$ indicates the symmetry plane.}
  \label{fig:modes}
\end{figure}

The combination of a neighboring sphere and two image systems
contribute
$b^e_{i\alpha,j\beta} = G^S_{\alpha\beta}(\vec r_i - \vec r_j)
  + G^W_{\alpha\beta}(\vec r_i - \vec R_i) +
  G^W_{\alpha\beta}(\vec r_i - \vec R_j)$
to the mobility of sphere $i$ in the $\alpha$ direction.
Eigenvectors of the corresponding diffusivity tensor appear in
Fig.~\ref{fig:modes}.
The independent modes of motion are rotated with respect to the
bounding wall by an amount which depends strongly on both
$r$ and $h$.
The experimentally measured in-plane motions
clearly are not independent yet
will satisfy Stokes-Einstein relations,
nonetheless, with pair-diffusion coefficients
$D^{C,R}_\alpha(r,h) = 
D_{1\alpha,1\alpha}(r,h) \pm D_{1\alpha,2\alpha}(r,h)$,
where the positive sign corresponds to collective motion, the negative to 
relative motion, and $\alpha$ indicates directions either perpendicular
or parallel to the line connecting the spheres' centers.
Explicitly, we obtain
\begin{eqnarray}
  \frac{D^{C,R}_\perp(r,h)}{D_0} & = & 1 - \frac{9}{16} \, \frac{a}{h} \pm
    \frac{3}{4} \, \frac{a}{r} \left[
      1 - \frac{1 + \frac{3}{2} \, \xi}{(1 + \xi)^{3/2}}
    \right] \\
  \frac{D^{C,R}_\|(r,h)}{D_0} & = & 1 - \frac{9}{16} \, \frac{a}{h} \pm 
    \frac{3}{2} \, \frac{a}{r} \left[
      1 - \frac{1 + \xi + \frac{3}{2} \, \xi^2}{(1 + \xi)^{5/2}} 
    \right]
\end{eqnarray}
up to ${\cal O} (a^3/r^3)$ and ${\cal O} (a^3/h^3)$, where 
$\xi = 4h^2/r^2$.
These results appear as solid curves in Fig.~\ref{fig:data}.

\begin{figure}[htbp]
\centering
    \includegraphics[width=3in]{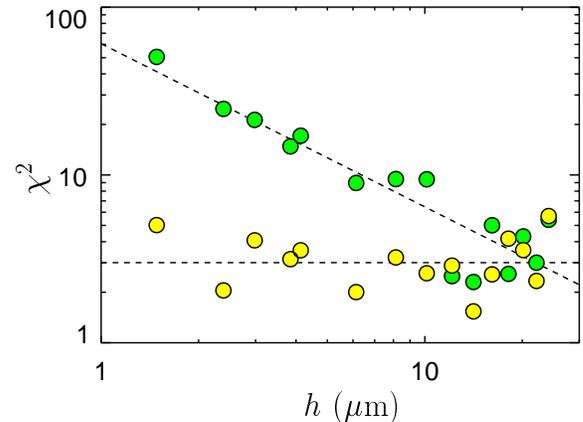}
    \caption{Mean-squared deviations between measured
      and predicted diffusion coefficients for relative
      perpendicular motion, averaged over initial
      separations $r$.
      Leading-order stokeslet analysis yields uniformly good predictions
      for all accessible values of $h$.
      Linear superposition, on the other hand, underestimates
      the wall's influence.
      Dashed lines are guides to
      the eye and emphasize the $a/h$ leading-order correction
      to the linear superposition model's prediction.}
    \label{fig:chisq}
\end{figure}

To gauge the success of this procedure and to quantify the range over
which the presence of a wall measurably influences colloidal
dynamics, we computed the error-weighted mean-squared
deviation of the predicted
diffusivities from the measured values,
$\chi_\psi^2(h) = \int \left[\left(D_\psi^{expt}(r,h) - D_\psi(r,h)\right)
/ \delta D_\psi^{expt}(r,h)\right]^2 \, dr$.
Typical results appear in Fig.~\ref{fig:chisq}.
The lowest-order stokeslet presented here 
analysis agrees well with measurement
over the entire experimentally accessible range.
Deviations from the linear superposition approximation's predictions,
on the other hand,
are evident out to $h = 15$~\micron or 30 radii.

The present study demonstrates that a confining surface can
influence colloidal dynamics over a large range of separations.
This influence is inherently a many-body effect, as
demonstrated by the
linear superposition approximation's failure.
Quantitative agreement between our measurements and a
leading-order stokeslet analysis offers hope for future
progress in understanding confinement's effects on colloidal
dynamics.

David Altman developed the transparent thin film heater with support
from the MRSEC REU program at the University of Chicago.
Work at the University of Chicago was supported by the National Science
Foundation through grant DMR-9730189, through the MRSEC program of
the NSF through grant DMR-9888595, and by the David and Lucile
Packard Foundation.  Theoretical work was supported by the A.P. Sloan 
Foundation,
the Mathematical Science Division of the National Science Foundation, and
a NDSEG Fellowship to TS.


\end{document}